\documentclass[journal]{IEEEtran}
\usepackage{amsmath,amssymb,amsfonts,array,tabularx,booktabs,cite,graphicx,tikz,xcolor}
\usepackage{soul,bbm,balance,etoolbox}
\newcommand{\compactdisplay}{\setlength{\abovedisplayskip}{3pt plus 2pt minus 1pt}\setlength{\belowdisplayskip}{3pt plus 2pt minus 1pt}\setlength{\abovedisplayshortskip}{2pt plus 1pt}\setlength{\belowdisplayshortskip}{3pt plus 1pt minus 1pt}}
\AtBeginEnvironment{equation}{\compactdisplay}
\AtBeginEnvironment{align}{\compactdisplay}
\DeclareRobustCommand{\revhl}[1]{#1}
\DeclareRobustCommand{\casehl}[1]{#1}
\DeclareRobustCommand{\newhl}[1]{#1}
\newcommand{\revmath}[1]{$\displaystyle #1$}
\usepackage[colorlinks=true,allcolors=blue,bookmarks=false]{hyperref}
\usetikzlibrary{arrows.meta,positioning,fit}

\newcommand{\NA}{\textemdash}
\newcommand{\indicator}{\mathbbm{1}}
\newcounter{algorithm}

\title{RF-Agent: Hierarchical Language-Agent Control for Instruction-Conditioned Active Spectrum Sensing}
\author{Hao~Zhang,~
Dongfang~Xu,~
Hang~Zou,~
Anis~Bara,
Brahim~Mefgouda,~\\
Lina~Bariah,~
Yuzhi~Yang,~
Fuhui~Zhou,~
Qihui~Wu,~
and M{\'e}rouane Debbah
\thanks{H. Zhang, H. Zou, A. Bara, B. Mefgouda, L. Bariah, and M. Debbah are with the Digital Future Institute, Khalifa University, Abu Dhabi, UAE. (e-mails: \{hao.zhang, hang.zou, anis.bara, brahim.mefgouda, lina.bariah, merouane.debbah\}@ku.ac.ae).}
\thanks{Y. Yang is with the School of Information Science, Japan Institute of Science and Technology (JAIST), Nomi 923-1292, Japan. (e-mail: yangyuzhi.zju@gmail.com)}
\thanks{D. Xu, F. Zhou, and Q. Wu are with the College of Artificial Intelligence, Nanjing
University of Aeronautics and Astronautics, Nanjing 211106, China, and also
with the Key Laboratory of Dynamic Cognitive System of Electromagnetic
Spectrum Space (Nanjing University of Aeronautics and Astronautics), Ministry
of Industry and Information Technology, Nanjing 211106, China. (e-mails: xudongfang@nuaa.edu.cn, zhoufuhui@ieee.org, wuqihui2014@sina.com)}}
\begin{document}
\flushbottom
\clubpenalty=1000
\widowpenalty=1000
\displaywidowpenalty=1000
\finalhyphendemerits=1000000
\bstctlcite{BSTcontrol}
\maketitle
\begin{abstract}
\revhl{Active radio-frequency (RF) sensing must acquire evidence under receiver limits while following confirmation, recovery, and stopping instructions. \newhl{We propose RF-Agent, a closed-loop architecture combining language supervision, episode memory, deterministic RF execution, perception feedback, and evidence-based reporting.} An independent auditor checks trajectory compliance. We derive acquisition and supervisory-request bounds, characterize full-history token complexity, and bound joint success by output validity and evidence capacity. \newhl{ActiveRF-AD tests instruction changes at fixed states and evaluates joint success, requiring compliance and RF correctness in the same episode.} \newhl{Training on different instructions at shared sensing states raises counterfactual-pair accuracy from 24.03\% to 96.22\% and Core joint success from 65.13\% to 72.12\%.} \newhl{Of the 6.99-point joint-success gain, 6.73 points reflect a decrease in the fraction of all episodes that are RF-correct but noncompliant.} \newhl{At Core budget three, paired-training hierarchical control achieves 66.60\% joint success versus 9.17\% for direct-action control.} \newhl{In a separate comparison at the same budget, RF-Agent improves joint success by 2.31 points and reduces RF acquisitions by 12.3\% relative to an initial-state planner, at increased inference cost.} Cross-backbone diagnostics cover three model families. These results show how instruction-responsive control improves task completion beyond what RF accuracy alone reveals.}
\end{abstract}

\section{Introduction}
\label{sec:introduction}
\IEEEPARstart{S}{pectrum} analysis provides information about radio-frequency (RF) activity that supports spectrum management and interference monitoring. Learning-based spectrum sensing and signal recognition extract such information by inferring occupancy, modulation, or anomalous activity from acquired signals \cite{oshea2016convolutional,zhang2026wsrsurvey}. Practical spectrum monitoring also requires deciding which observations to acquire and when to stop. Limited instantaneous receiver bandwidth restricts the frequency regions observed in each acquisition, while a finite sensing budget limits the number of measurements. The order of acquisitions can therefore affect whether relevant evidence becomes available before the budget is exhausted. Evidence requirements also vary with the task: an interference investigation may demand repeated confirmation, whereas a routine occupancy check may prioritize releasing the \mbox{receiver early.}

Active perception treats observation acquisition as task dependent \cite{bajcsy1988active}, while cognitive radio and dynamic spectrum access connect sensing with adaptive decisions \cite{haykin2005cognitive,akyildiz2006next}. When monitoring requirements are expressed in natural language, a controller must interpret confirmation requirements, failure-handling rules, and stopping conditions and apply them to the evolving sensing state. Consequently, identical observations and remaining resources can call for different actions under different instructions. For example, following an acquisition timeout with budget and receiver views still available, a recovery instruction requires another measurement, whereas an abort instruction \mbox{requires termination.}

Language agents connect natural-language task descriptions with execution feedback to select domain tools \cite{yao2023react,tong2025wirelessagent}. \revhl{The central challenge is to translate instruction-dependent evidence requirements into executable acquisition and stopping decisions, and to verify their fulfillment separately from RF detection accuracy. RF-Agent implements a single-agent sensing loop: given an instruction, it retains execution history, selects acquisition strategies, invokes receiver tools, and updates decisions from feedback. The \newhl{large language model (LLM)} implements supervision; explicit components handle perception, state, resource accounting, \mbox{and reporting.}}

\revhl{A correct RF report can still result from a trajectory that violates confirmation, recovery, or stopping requirements. We therefore evaluate joint success, requiring compliance and RF correctness in the same episode. Matched-state instruction tests and complete-trajectory audits connect supervisory decisions to task completion; interface and planning comparisons examine execution reliability and acquisition cost.} \revhl{ActiveRF-AD provides synthetic RF scenes and receiver views under shared budgets, perception, and reporting. Supervisors see instructions and acquired evidence, excluding hidden truth and unselected observations. The reporter and auditor use evaluator-provided task specifications.} Our contributions are:
\begin{enumerate}
\item \revhl{We develop RF-Agent, a hierarchical architecture that connects instruction-conditioned strategy selection to executable RF acquisition and feedback-driven stopping. Shared episode state, receiver execution, and reporting support both learned language and lightweight supervisory policies.}
\item \revhl{We characterize acquisition and supervisory-request bounds, full-history context complexity, and a joint-success upper bound combining output validity with evidence capacity. These properties explain how execution errors, evidence limits, and repeated supervision constrain task success and cost.}
\item We develop ActiveRF-AD and an evaluation protocol combining matched-state instruction interventions with trajectory audits based on task requirements. The protocol distinguishes reference-action agreement from task compliance, accepting compliant trajectories whose actions differ from those prescribed by a reference policy.
\item \revhl{We quantify how supervisory training and control design affect task completion. Hard-state training raises Core joint success by 6.99 percentage points, including \newhl{a 6.73-point decrease in the fraction of all episodes that are RF-correct but noncompliant.} Paired-training interface and independent planning comparisons establish execution-reliability gains and a budget-dependent acquisition--inference tradeoff.}
\end{enumerate}

The remainder of this paper is organized as follows. Section~\ref{sec:related} reviews related work. Section~\ref{sec:task} formulates the instruction-conditioned sensing task, and Section~\ref{sec:framework} presents the RF-Agent framework. Section~\ref{sec:experiments} presents the experimental setup, evaluation protocol, and results. Finally, Section~\ref{sec:conclusion} concludes the paper.

\section{Related Work}
\label{sec:related}
\subsection{Active Sensing and Cognitive Radio}
Active perception and partially observable control treat observation selection as part of task execution \cite{bajcsy1988active,kaelbling1998planning}. Cognitive radio connects sensing with adaptive spectrum use \cite{haykin2005cognitive,akyildiz2006next}. Spectrum-sensing studies address reliability, duration, throughput, fading, and noise uncertainty \cite{axell2012spectrum,liang2008sensing}. Reinforcement learning supports spectrum-access and resource-management policies \cite{naparstek2019dsa,luong2019drl}; RFRL Gym provides an environment for interactive cognitive-radio experiments \cite{rosen2023rfrlgym}. 
For sequential wideband sensing, Wu \textit{et al.}~\cite{wu2023wideband} jointly address channel selection and stopping through a Markov decision process with sampling costs, sensing requirements, and a finite budget. Their model-augmented deep reinforcement learning minimizes sampling and detection-decision costs. Our task treats instruction-specific confirmation, recovery, and stopping as trajectory obligations, while evaluating RF correctness and acquisition \mbox{cost separately.}

\subsection{RF Perception and Representation Learning}
RF perception uses statistical features \cite{dobre2007survey} and learned representations \cite{oshea2016convolutional,zhang2026wsrsurvey}, including over-the-air and distributed sensing under channel and deployment constraints \cite{oshea2018overair,rajendran2018deep}. Modulation classifiers address multiple scales \cite{zhang2022multiscale}, limited labels with SSwsrNet \cite{zhang2024sswsrnet}, and few-shot open-set recognition over multipath fading with FSOS-AMC \cite{zhang2025fsosamc}. SpectrumFM reuses representations across classification, sensing, and anomaly detection \cite{zhou2026spectrumfm}; spectrum cognition connects signal analysis to semantic situation understanding \cite{zhang2026spectrumcognition}. RF-Agent fixes perception across controllers to study how its outputs, instructions, and execution history guide acquisition \mbox{and stopping.}

\subsection{Language Agents for Wireless and Signal Processing}
Telecom language research spans document understanding \cite{bariah2023telecom}, domain adaptation in TelecomGPT \cite{zou2024telecomgpt}, and retrieval support for compact models \cite{piovesan2024telecom}. RF-GPT connects RF observations with language by encoding spectrograms as RF tokens and instruction-tuning a multimodal language model on synthetic RF--text data \cite{zou2026rfgpt}. It supports questions, explanations, and structured outputs about acquired \mbox{RF scenes.}

Language agents extend interpretation to tool-mediated execution. ReAct interleaves reasoning with actions and external feedback \cite{yao2023react}, tool-learning methods invoke specialized operations \cite{schick2023toolformer}, and Reflexion uses verbal feedback to improve subsequent attempts \cite{shinn2023reflexion}. \revhl{WirelessAgent organizes perception, memory, planning, and action for network management, demonstrated through feedback-driven resource allocation in network slicing}~\cite{tong2025wirelessagent}\revhl{. SignalLLM automates signal-processing workflows through task decomposition, hierarchical planning, solution refinement, and execution by reasoning, code synthesis, or model invocation}~\cite{ke2025signalllm}\revhl{. \newhl{RF-Agent controls which receiver view to acquire next and when to stop under instruction-specific evidence and recovery requirements. Its constrained language interface links these decisions to deterministic RF execution; independent audits check evidence, recovery, and stopping obligations. Compliance, RF correctness, and acquisition cost are evaluated over the same sensing trajectories.}} Ferrag \textit{et al.}~\cite{ferrag2026agents} propose an agentic network architecture that places semantic abstraction and hierarchical reasoning above deterministic \mbox{network infrastructure.}

Temporal-logic approaches connect language instructions to constrained execution. Safety Chip translates safety constraints into human-verified linear temporal logic and monitors proposed robot actions \cite{yang2024safetychip}; LIMP compiles instructions into automata for task and motion planning \cite{quartey2025limp}. Agent-C enforces temporal and state-dependent tool policies during generation \cite{kamath2025agentc}. RF-Agent audits sensing trajectories after execution, checking instruction-dependent evidence requirements and continuation obligations under finite \mbox{acquisition budgets.}

\subsection{Evaluating Wireless-Agent Behavior}
Stateful agent benchmarks assess tool use through executable criteria. ToolSandbox checks intermediate and final milestones along interactive trajectories \cite{lu2025toolsandbox}, while $\tau$-bench compares final database states with task goals after agent--user interactions under domain policies \cite{yao2025taubench}. WirelessBench evaluates wireless reasoning, resource allocation, and proactive decisions using numerical tolerances, quality-of-service checks, and diagnostic categories \cite{tong2026wirelessbench}. 
ActiveRF-AD evaluates whether sensing instructions require or prohibit further acquisition under the available resources. Its trajectory audits accept compliant alternatives to teacher actions and assess compliance separately from \mbox{RF correctness.}

\section{Instruction-Conditioned Sensing Task}
\label{sec:task}
We consider receive-only anomaly detection under natural-language instructions and a finite acquisition budget. The controller selects acquisitions and termination; a shared reporter applies task-specific evidence requirements. The objective is a correct scene-level report reached through a compliant \mbox{sensing trajectory.}

\subsection{Sensing State and Actions}
\revhl{Let \mbox{$p_{\mathcal Z}$} be the scene distribution and \mbox{$z\sim p_{\mathcal Z}$} a hidden RF scene} with label $y(z)\in\{\mathrm{anomaly},\mathrm{normal}\}$, and let $x$ denote the instruction. \revhl{The catalog \mbox{$\mathcal V=\{v_1,\ldots,v_N\}$} contains \mbox{$N$} \newhl{receiver configurations (views), each specifying center frequency and bandwidth.}} The effective budget is $B=\min(B_{\mathrm{requested}},N)$, where $B_{\mathrm{requested}}$ is a positive integer. Before decision $t$, the public state is
\begin{equation}
s_t=(B_t,\mathcal V_t,h_t,\mathcal O_t,\mathcal E_t),
\label{eq:state}
\end{equation}
where $B_t$ and $\mathcal V_t$ are the remaining budget and views, $h_t$ records interface-specific action labels, and $\mathcal O_t$ and $\mathcal E_t$ record observations and failures in order. Initially, $B_1=B$, $\mathcal V_1=\mathcal V$, and histories are empty; the initial budget and catalog geometry remain public. A successful observation exposes the selected view and its detector decision, score, confidence, and time--frequency boxes; a failure exposes its view, attempt index, type, and recoverability. Scene truth, generation metadata, and unselected \newhl{in-phase/quadrature (IQ) samples} and perception outputs remain hidden.

While budget and views remain available, the executable actions are
\begin{equation}
a_t\in\{\textsc{Acquire}(v):v\in\mathcal V_t\}\cup\{\textsc{Stop}\}.
\label{eq:receiver-action}
\end{equation}
An attempt at view $v_t$, whether successful or failed, updates
\begin{equation}
B_{t+1}=B_t-1,\qquad
\mathcal V_{t+1}=\mathcal V_t\setminus\{v_t\}.
\label{eq:transition}
\end{equation}
Success appends a perception record to $\mathcal O_t$; failure appends to $\mathcal E_t$ without supplying RF evidence. Each decision updates $h_t$. \textsc{Stop} terminates without acquisition, and budget or view exhaustion terminates automatically. The trajectory $\tau$ records states, decisions, attempted views, outcomes, and termination.

\subsection{Instruction-Dependent Requirements}
\revhl{Each benchmark instruction \mbox{$x$} is paired with a reference specification \mbox{$g=(k_+,k_-,f,L,d_-,d_+)$} defining its hard obligations. The evaluator obtains \mbox{$g$} from the predefined task configuration, independently of the supervisor's interpretation of \mbox{$x$} and of hidden RF scene truth.} Task profiles additionally express preferences over RF outcomes and acquisition cost; prescribed teacher policies select specific actions that reflect those preferences and policy conventions. Positive integers $k_+$ and $k_-$ set the anomaly and normal evidence thresholds. The failure rule $f$ specifies no additional obligation, retry after a recoverable failure, or abort after any failure. The optional nonnegative integer $L$ caps attempts before a positive observation; binary flags $d_-$ and $d_+$ govern mandatory stopping and continuation. Ordinary controllers receive $x$ and public state; only the shared reporter, auditor, and designated specification-aware references receive $g$.

\revhl{Let \mbox{$p_t,n_t$} count positive and normal outputs, and \mbox{$A_t=B-B_t$} count attempts. A further acquisition is required while permitted if \mbox{$0<p_t<k_+$}, or if \mbox{$p_t=0,n_t<k_-$}. Under retry, a recoverable latest failure requires a further permitted attempt; repeated failures renew this obligation. Table}~\ref{tab:requirement-examples} \revhl{illustrates how evidence and task requirements determine control \mbox{and reporting.}}

\revhl{The flag \mbox{$d_+=1$} expresses continued search after a detection, as specified by the budgeted-search instruction. It can therefore require extra acquisitions after anomaly evidence is sufficient; their cost is recorded in the call count and utility. The cap \mbox{$L$} counts all acquisition attempts, including failures, imposes no acquisition minimum, and ceases to apply after a positive. It takes precedence over retry. Abort requires stopping after any failure; \mbox{$d_-=1$} requires stopping once \mbox{$p_t=0,n_t\geq k_-$}.}

\revhl{The executor enforces budget and view availability; caps, abort, and mandatory stopping are audited controller obligations. A continuation requirement applies only while its next acquisition is permitted, even if the remaining resources cannot support a binary report. Thus, resource-limited evidence and stopping despite a feasible obligation \mbox{are distinguished.}}

\begin{table}[t]
\centering
\caption{\revhl{Evidence, Control Obligations, and Reports}}
\label{tab:requirement-examples}
\scriptsize
\setlength{\tabcolsep}{3pt}
\renewcommand{\arraystretch}{1.12}
\begin{tabularx}{\columnwidth}{@{}>{\raggedright\arraybackslash}X>{\raggedright\arraybackslash}p{0.21\columnwidth}>{\raggedright\arraybackslash}p{0.20\columnwidth}@{}}
\toprule
\revhl{Current state and requirement} & \revhl{Control obligation} & \revhl{Report if ended}\\
\midrule
\revhl{\mbox{$p_t=1,n_t=2$}; \mbox{$k_+=k_-=2$}} & \revhl{Acquire if permitted} & \revhl{Abstain}\\
\revhl{\mbox{$p_t\geq k_+$}; \mbox{$d_+=1$}} & \revhl{Acquire if permitted} & \revhl{Anomaly}\\
\revhl{\mbox{$p_t=0,n_t\geq k_-$}; \mbox{$d_-=1$}} & \revhl{Stop} & \revhl{Normal}\\
\revhl{Latest failure recoverable; retry} & \revhl{Acquire if permitted} & \newhl{Apply reporting rule}\\
\revhl{\mbox{$p_t=0,A_t\geq L$}, or failure with abort} & \revhl{Stop} & \newhl{Apply reporting rule}\\
\bottomrule
\end{tabularx}
\par\vspace{2pt}
\begin{minipage}{\columnwidth}
\scriptsize\raggedright
\revhl{Rows illustrate separate cases with no other active stopping obligation in acquisition rows. ``If permitted'' requires available budget and views and no applicable cap or abort. Resource exhaustion terminates automatically. Reports use}~\eqref{eq:report}\revhl{; a supported report alone does not certify compliant termination.}
\end{minipage}
\end{table}

\subsection{Reporting and Task Success}
After a syntactically valid \textsc{Stop} or automatic resource termination, the shared reporter uses the final positive and normal counts $p(\tau)$ and $n(\tau)$:
\begin{equation}
\hat y(\tau;g)=
\begin{cases}
\mathrm{anomaly}, & p(\tau)\geq k_+(g),\\
\mathrm{normal}, & p(\tau)=0,\ n(\tau)\geq k_-(g),\\
\mathrm{abstain}, & \text{otherwise}.
\end{cases}
\label{eq:report}
\end{equation}
\revhl{Positive and normal counts summarize distinct frequency views of the same time interval, not mutually exclusive votes on a repeatedly resampled scene. A normal view elsewhere therefore does not cancel a positive detector output: once \mbox{$p\geq k_+$}, later normal outputs cannot reverse the anomaly report. A normal report requires \mbox{$p=0$} so that unresolved positive evidence cannot be outvoted. For \mbox{$k_+=k_-=2$}, even \mbox{$(p,n)=(1,5)$} yields abstention: the positive lacks confirmation and prevents a normal claim. Reporting retains evidence on abort and is separate from termination compliance. Views can be correlated; the thresholds express task requirements rather than independent statistical confidence. Abstention for insufficient evidence is distinct from open-set recognition}~\cite{scheirer2013open}\revhl{.}

\revhl{The reporter uses only the evidence thresholds \mbox{$(k_+,k_-)$} and acquired detector outputs, without access to hidden scene truth. The supervisor interprets \mbox{$x$} to select acquisitions and stopping, while the auditor checks the resulting trajectory against the full reference \mbox{$g$}. The evaluation supplies trusted task requirements to the reporter and withholds them from the supervisor to isolate instruction-conditioned control. Deployment can obtain \mbox{$g$} from operator-configured templates or structured task fields consistent with \mbox{$x$}. Language-only requests additionally require instruction-to-specification extraction and semantic validation, such as operator confirmation. The reported evaluation begins with a verified \mbox{task specification.}}

For the rule checks $\mathcal K(g,\tau)$ triggered along the trajectory, including final report support, compliance is
\begin{equation}
C(\tau;g)=
\prod_{k\in\mathcal K(g,\tau)}c_k(\tau;g),
\qquad c_k\in\{0,1\},
\label{eq:compliance}
\end{equation}
where $c_k=1$ denotes no violation. A continuation obligation is violated by stopping only when another call is permitted by the constraints above. Abstention makes no binary claim, so it passes the report-support check. Other violations still result in noncompliance. The audit in Section~\ref{sec:obligation-audit} evaluates compliance independently of teacher-action agreement and \mbox{RF correctness.}

\revhl{Let \mbox{$\indicator[\cdot]$} equal one when its condition holds and zero otherwise. For episode \mbox{$e$}, \mbox{$R_e=\indicator[\hat y(\tau_e;g_e)=y(z_e)]$} indicates RF correctness. Joint success over \mbox{$N_{\mathrm{ep}}$} episodes is}
\begin{equation}
\text{\revmath{J=\frac{1}{N_{\mathrm{ep}}}\sum_{e=1}^{N_{\mathrm{ep}}}C(\tau_e;g_e)R_e.}}
\label{eq:joint-success}
\end{equation}
Thus, success requires compliance and RF correctness in the same episode. Abstentions and controller errors receive no correctness credit and remain in the denominator. RF preferences, acquisition costs, and inference costs are evaluated separately in Section~\ref{sec:outcome-metrics}.

\section{\revhl{RF-Agent Architecture}}
\label{sec:framework}
\begin{figure*}[t]
\centering
\includegraphics[width=170mm]{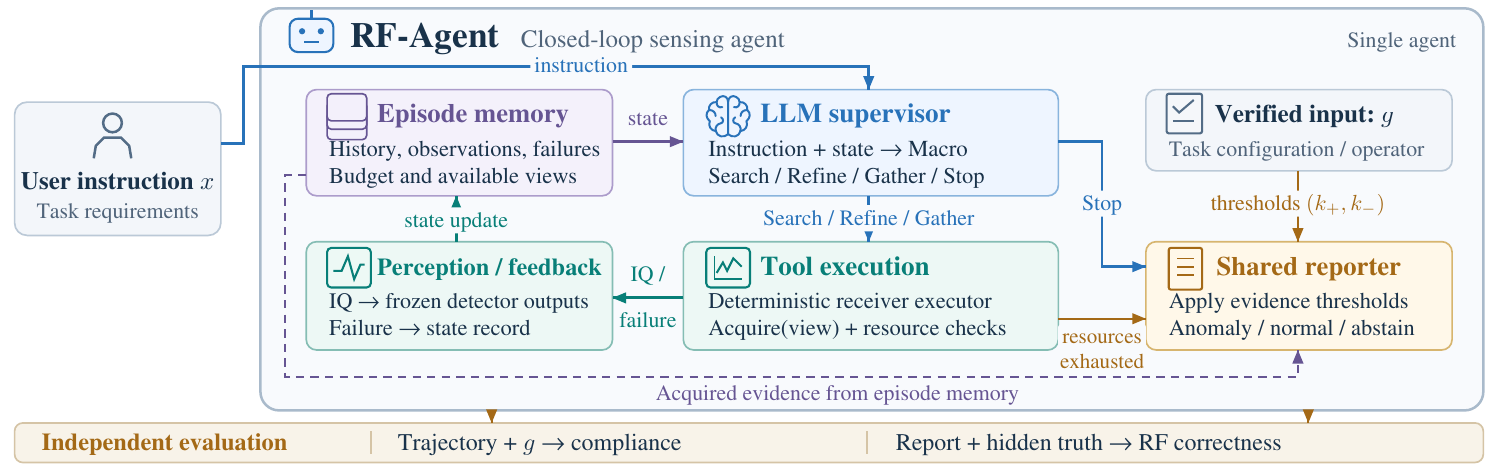}
\caption{\revhl{RF-Agent architecture and feedback loop. Dashed arrows denote access to stored evidence; evaluation is separate from online control.}}
\label{fig:architecture}
\end{figure*}

\revhl{RF-Agent combines instruction-conditioned supervision, episode memory \mbox{$s_t$}, receiver tools, perception feedback, and reporting (Fig.}~\ref{fig:architecture}\revhl{). The supervisor selects Macros, tool feedback updates memory, and Stop or resource exhaustion triggers reporting. An independent auditor evaluates \mbox{completed trajectories.}}

\subsection{Instruction-Conditioned Supervision}
\label{sec:execution}
\revhl{At each nonterminal decision, the LLM implements the supervisory policy \mbox{$\pi$}, mapping the instruction and updated public state, including catalog metadata and initial budget, to a strategy label (Macro):}
\begin{align}
m_t&=\pi(x,s_t),\nonumber\\
m_t&\in\{\textsc{Search},\textsc{Refine},\textsc{Gather},\textsc{Stop}\}.
\label{eq:macro-action}
\end{align}
\textsc{Search} requests coverage expansion, \textsc{Refine} requests acquisition guided by a detected cue, and \textsc{Gather} requests additional evidence, including recovery attempts. \textsc{Stop} requests termination and reporting. The history $h_t$ stores preceding Macro labels. The supervisor selects these labels using public evidence and failure feedback; it receives neither scene truth nor evaluator-provided $g$. Section~\ref{sec:control-decomposition} describes the corresponding inputs for direct-action comparisons.

\begin{figure}[t]
\centering
\includegraphics[width=\columnwidth]{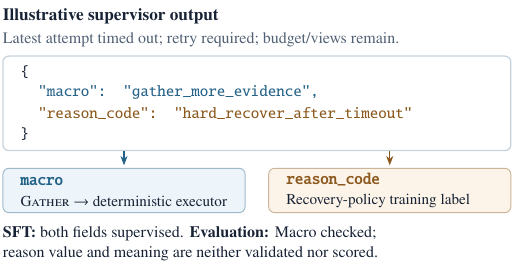}
\caption{\revhl{Example supervisor JSON for timeout recovery.}}
\label{fig:supervisor-json}
\end{figure}

\revhl{The supervisor returns an action message with exactly two JSON string fields: \mbox{\texttt{macro}} and \mbox{\texttt{reason\_code}}. Fig.}~\ref{fig:supervisor-json} \revhl{illustrates a recovery decision. The Macro drives execution, while the auxiliary reason code provides action-associated \mbox{training supervision.}}

\revhl{\newhl{Supervised fine-tuning (SFT)} supervises both fields; the reason-code vocabulary is provided in the supplementary material. Evaluation checks the response envelope, JSON object, and Macro membership. The \newhl{strategy-campaign Macro parser} ignores the reason field; the control-interface and planning comparisons additionally require exactly the two string fields above for hierarchical outputs. Neither parser validates reason-code vocabulary membership or semantic consistency, nor is reason content separately scored. After structural checks, only the Macro determines the controller action.} Formal evaluation disables policy fallback. In matched-state evaluation, an invalid response counts as an incorrect action. In closed-loop evaluation, it ends the episode with abstention and noncompliance, without an additional RF call. The episode retains all previously incurred costs and remains in all episode denominators. Transport or service failures pause \mbox{resumable evaluation.}

\subsection{Deterministic Execution and Perception Feedback}
\label{sec:receiver-execution}
For a nonterminal Macro, the executor selects $v_t=\pi_{\mathrm{low}}(s_t,m_t)\in\mathcal V_t$ and performs $a_t=\textsc{Acquire}(v_t)$. It enforces total budget and view availability; spectral coverage changes only after successful acquisitions. Coverage is the union of successfully observed frequency intervals. \textsc{Search} maximizes uncovered bandwidth, breaking ties by proximity to the midpoint of the catalog's frequency extent, then lower center frequency and \mbox{view index.}

For cue-guided acquisition, observations are ranked by positive decision, then anomaly score, retaining the earlier observation for ties. The selected observation's localization box supplies a target frequency. Both \textsc{Refine} and \textsc{Gather} select the available view nearest this target, with view-index tie breaking; missing or nonfinite boxes invoke the coverage rule. These labels distinguish prescribed supervisory intentions but produce the same receiver action at a fixed state. Their agreement with teacher labels therefore measures a different property from the resulting acquisition and \mbox{stopping behavior.}

\revhl{A frozen spectral-residual detector maps each acquired view to a decision, score, confidence, and estimated time--frequency box. The environment appends these outputs or a failure record to episode state before the next supervisory call. Thus, the agent acquires new evidence through tools; the LLM receives structured feedback rather than raw IQ.} Detector configuration and calibration are specified in Section~\ref{sec:perception-setup}. \textsc{Stop} or resource exhaustion invokes the shared reporter in \eqref{eq:report}. Its access to $g$ governs reporting only; instruction-specific continuation and stopping remain supervisory decisions evaluated by \mbox{the auditor.}

\subsection{\revhl{Supervisory Learning}}
\label{sec:supervisor-learning}
\revhl{The teacher supplies \mbox{$m^\star(x,s_t)=\pi_{\mathrm T}(s_t;\rho_x)$} using a prescribed policy identity \mbox{$\rho_x$}. This identity maps task preferences and policy conventions to unique reference labels. The trajectory auditor instead checks hard task obligations, so teacher-action agreement and trajectory compliance measure \mbox{distinct properties.}}

\revhl{\newhl{The balanced, recall-priority, and efficient profiles share the same hard obligations in \mbox{$g$} but use different stopping conventions. For the original two-positive confirmation profile, the teacher selects Gather when the latest attempt fails and budget remains; this retry convention adds no hard obligation.} The supplementary material gives the teacher's policy priorities. The contract reference instead uses \mbox{$(g,s_t)$} to select admissible actions.} \revhl{The contract policy prioritizes resource and stopping constraints, then retry and positive continuation; it searches initially and gathers additional evidence. Both specification-aware policies provide \mbox{execution references.}}

\revhl{Training adapts only the supervisory policy; perception, state updates, and receiver execution remain fixed. \newhl{Standard SFT} learns acquisition and stopping choices from demonstrations \mbox{\cite{ouyang2022instruct}}. Hard-state SFT pairs different instructions with shared public states, including partial observations and failures, to train instruction-dependent choices at fixed RF evidence. Prompt tokens are loss-masked; supervision covers the Macro and \mbox{reason code.}}

\revhl{We use direct preference optimization (DPO) to study how RF-outcome preferences affect instruction-conditioned control \mbox{\cite{rafailov2023dpo}}. Preferences rank candidate first actions followed by teacher rollouts but omit audited compliance. This diagnostic tests alignment between RF-proxy optimization and trajectory compliance. Training configurations and preference construction appear in Section}~\ref{sec:controller-setup}.

\subsection{Independent Trajectory Auditing}
\label{sec:obligation-audit}
\revhl{The auditor independently checks each recorded decision against the specification, action/view ledger, and accumulated evidence. It records violations, feasible continuation exposures, and resource-blocked obligations; the full pseudocode is in the supplement. A premature stop violates an acquisition obligation only if another call is permitted. Resource limitation flags blocked evidence or retry obligations at termination; budget exhaustion required solely by \mbox{$d_+$} does \mbox{not qualify.}}

\revhl{This separates admissible behavior from teacher agreement: after one positive under two-positive confirmation, any permitted acquisition satisfies continuation, even if its Macro differs from the teacher's choice. Conversely, a failed first attempt with a one-call budget yields compliant, resource-limited abstention under retry. Measurement provenance is an external input; RF correctness uses hidden truth separately \mbox{from auditing.}}

\subsection{\revhl{Execution Bounds and Success Feasibility}}
\label{sec:theory-analysis}
\newhl{The following bounds connect resource limits to supervisory cost and separate output reliability from the availability of sufficient evidence.}
\paragraph{\revhl{Proposition 1: bounded execution}}
\revhl{Let \mbox{$A_e$} count RF attempts and \mbox{$Q_e$} completed per-step supervisory decisions in episode \mbox{$e$} with budget \mbox{$B_e$}; service retries are excluded. Every nonterminal decision executes one acquisition; Stop or an output error terminates, and exhaustion is automatic. Therefore,}
\begin{equation}
\text{\revmath{Q_e=A_e+\mathbbm{1}[A_e<B_e]\leq B_e.}}
\label{eq:request-bound}
\end{equation}
\revhl{If fixed prompt content uses at most \mbox{$L_{\mathrm{base}}$} tokens and each attempt record adds at most \mbox{$h_{\mathrm{tok}}$}, episode input-token \mbox{volume satisfies}}
\begin{equation}
\text{\revmath{T_{\mathrm{in},e}\leq Q_eL_{\mathrm{base}}+\frac{h_{\mathrm{tok}}Q_e(Q_e-1)}{2}.}}
\label{eq:input-token-bound}
\end{equation}
\revhl{Let \mbox{$T_{\mathrm{out},e}$} denote the total output-token count in episode \mbox{$e$}. An output cap \mbox{$L_{\mathrm{out}}$} per request gives \mbox{$T_{\mathrm{out},e}\leq Q_eL_{\mathrm{out}}$}. Thus, requests grow at most linearly with budget, whereas retransmitted history can produce quadratic input-token volume.}

\paragraph{\revhl{Proposition 2: joint-success envelope}}
\revhl{Let \mbox{$\varepsilon_e\in\{0,1\}$} equal one if a controller-output error occurs in episode \mbox{$e$} and zero otherwise, and let \mbox{$u_e$} upper-bound successful acquisitions in any compliant execution. Set \mbox{$\varphi_e=1$} when \mbox{$u_e\geq k_+(g_e)$} for an anomalous scene or \mbox{$u_e\geq k_-(g_e)$} for a normal scene, and zero otherwise. With bars denoting episode means,}
\begin{equation}
\text{\revmath{J\leq\frac{1}{N_{\mathrm{ep}}}\sum_e(1-\varepsilon_e)\varphi_e
\leq\min(1-\bar\varepsilon,\bar\varphi).}}
\label{eq:success-envelope}
\end{equation}
\revhl{Joint success requires valid output and enough successful observations for the correct report. This separates interface reliability from evidence feasibility without assuming independent observations. The supplement proves both propositions. Section}~\ref{sec:closed-loop-protocol} \revhl{applies the capacity bound to \mbox{changed requirements.}}

\paragraph{\revhl{Corollary: unavoidable abstention}}
\revhl{Let \mbox{$u_t$} bound additional successful acquisitions over all admissible continuations from \mbox{$s_t$}. If}
\begin{equation}
\text{\revmath{p_t+u_t<k_+,\quad (p_t>0\ \text{or}\ n_t+u_t<k_-),}}
\label{eq:state-abstention}
\end{equation}
\revhl{every such continuation ending through the shared reporter yields abstention. Indeed, the first inequality prevents anomaly confirmation; a normal report is excluded by persistent positives or insufficient normal-evidence capacity. For example, \mbox{$p_t=1$}, \mbox{$k_+=3$}, and \mbox{$u_t=1$} exclude both reports. This analytical certificate does not implement early stopping: instructions may still require a \mbox{permitted acquisition.}}

\section{Experiments}
\label{sec:experiments}
The evaluation follows three levels: agreement with prescribed supervisory choices, compliance and RF correctness over trajectories, and task-success and cost differences between controller designs. Table~\ref{tab:campaigns} summarizes four experiments sharing perception, receiver views, and reporting rules. \newhl{Results are reported separately for each campaign; shared scenes, adapters, and campaign-specific settings are identified in Tables}~\ref{tab:campaigns} \newhl{and}~\ref{tab:cohorts}.

\begin{table*}[t]
\centering
\caption{\revhl{Qwen3-8B Evaluation Campaigns}}
\label{tab:campaigns}
\small
\setlength{\tabcolsep}{4pt}
\renewcommand{\arraystretch}{1.12}
\begin{tabularx}{\textwidth}{@{}
  >{\raggedright\arraybackslash}p{0.17\textwidth}
  >{\raggedright\arraybackslash}p{0.25\textwidth}
  >{\raggedright\arraybackslash}p{0.22\textwidth}
  >{\raggedright\arraybackslash}X
  >{\centering\arraybackslash}p{0.065\textwidth}@{}}
\toprule
\textbf{Experiment} & \textbf{Model training} & \textbf{Output interface} & \textbf{Evaluation focus} & \textbf{Tokens}\\
\midrule
Strategy and\newline compliance & Standard SFT, hard SFT,\newline and hard DPO & Per-step Macro & Strategy agreement,\newline compliance, RF outcomes & 96\\
\addlinespace[4pt]
Instruction parsing & Same Macro adapters;\newline no parser SFT & Per-step Macro or\newline once-per-episode contract & Task completion and\newline inference cost & 256\\
\addlinespace[4pt]
Hierarchical/flat & New adapters; 20,000\newline paired states per model & Macro or direct\newline receiver action & Execution reliability under paired interface training & 256\\
\addlinespace[4pt]
Feedback/planning & New hierarchical adapter;\newline independently trained planner & Per-step Macro or\newline initial view plan & Joint success, RF acquisitions, and inference cost & 256\\
\bottomrule
\end{tabularx}
\par\vspace{3pt}
\begin{minipage}{\textwidth}
\footnotesize\raggedright
\textit{Note:} Tokens denote the output limit per request. The interface and planning comparisons use independent hierarchical runs on the same scenes.
\end{minipage}
\end{table*}

\subsection{Experimental Setup}
\label{sec:experimental-setup}
\subsubsection{ActiveRF-AD and Frozen Perception}
\label{sec:perception-setup}
ActiveRF-AD contains 24,300 deterministic synthetic IQ scenes and 145,800 \newhl{precomputed receiver views} (Table~\ref{tab:data}). Each scene combines a full-duration normal signal generated with TorchSig 2.1.1 \cite{boegner2022torchsig,boegner2022wideband} and noise, with an anomaly in half of each split. \revhl{At sample index \mbox{$n$}, the received signal \mbox{$r[n]$} combines normal and anomaly waveforms \mbox{$s[n;\theta_s]$} and \mbox{$a[n;\theta_a]$}, with parameters \mbox{$\theta_s,\theta_a$}, and independent circular complex Gaussian noise \mbox{$w[n]$}:}
\begin{equation}
\text{\revmath{r[n]=s[n;\theta_s]+\indicator[y(z)=\mathrm{anomaly}]a[n;\theta_a]+w[n].}}
\label{eq:scene}
\end{equation}
Normal-signal mean power is one, and noise power is $10^{-\mathrm{SNR}/10}$ for \newhl{signal-to-noise ratio (SNR)} drawn from 5--20 dB before receiver filtering. Normal bandwidth is 250--400 kHz, centered in $[-1.44,1.44]$ MHz. \newhl{The interference-to-signal ratio (ISR)} is the anomaly-to-normal mean-power ratio during the anomaly's active interval, excluding noise.

\begin{table}[t]
\centering
\caption{\revhl{ActiveRF-AD Dataset}}
\label{tab:data}
\small
\setlength{\tabcolsep}{4pt}
\renewcommand{\arraystretch}{1.12}
\begin{tabular*}{\columnwidth}{@{\extracolsep{\fill}}lrcl@{}}
\toprule
\textbf{Split} & \textbf{Scenes} & \textbf{ISR (dB)} & \textbf{Anomaly types}\\
\midrule
Training & 18,000 & $[-10,10]$ & Tone/chirp/pulse\\
\addlinespace[2pt]
Validation & 1,800 & $[-10,10]$ & Tone/chirp/pulse\\
\addlinespace[2pt]
Core & 1,200 & $[-10,10]$ & Tone/chirp/pulse\\
\addlinespace[2pt]
Stress & 2,100 & $[-20,-8]$ & Tone/chirp/pulse\\
\addlinespace[2pt]
OOD & 1,200 & $[-15,5]$ & Noise/hopping\\
\bottomrule
\end{tabular*}
\par\vspace{3pt}
\begin{minipage}{\columnwidth}
\footnotesize\raggedright
\textit{Note:} \revhl{Each scene has six 2-MHz views. Formal experiments use the fixed Core cohorts in Table}~\ref{tab:cohorts}\revhl{, holding the RF distribution and detector constant while varying instructions, interfaces, and feedback. Stress and OOD document additional \newhl{precomputed} dataset coverage; they are outside the present control comparison.}
\end{minipage}
\end{table}

\begin{figure*}[!t]
\centering
\includegraphics[width=\textwidth]{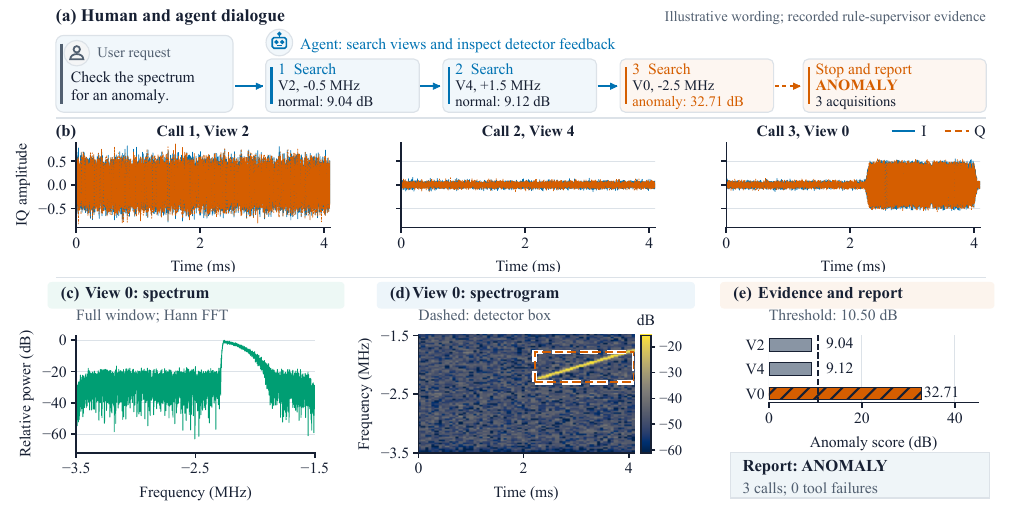}
\caption{\revhl{Illustrative user--agent interaction based on an archived rule-supervisor trace, with synthetic-IQ evidence. The illustrative 10.50-dB threshold differs from formal evaluation (10.34375 dB).}}
\label{fig:iq-case}
\end{figure*}

\revhl{For view \mbox{$i$}, \mbox{$f_i$} is its center frequency, \mbox{$h_i$} its filter, and \mbox{$F_s=8$} MS/s the input sample rate. Here \mbox{$j^2=-1$} and \mbox{$*$} denotes convolution. The operator \mbox{$\mathcal D_4$} resamples to 2 MS/s, and \mbox{$\mathcal C_{N_{\mathrm{samp}}}$} center-crops to \mbox{$N_{\mathrm{samp}}=8192$} samples. The cropped signal \mbox{$u_i$} gives the acquired IQ view \mbox{$\widetilde r_i$} after gain and componentwise clipping to \mbox{$[-4,4]$}:}
\begin{align}
u_i&=\text{\revmath{\mathcal C_{N_{\mathrm{samp}}}}}\left\{\mathcal D_4
\left[h_i*(r[n]e^{-j2\pi f_i n/F_s})\right]\right\},\nonumber\\
\text{\revmath{\widetilde r_i}}&=\operatorname{clip}_4(10^{-6/20}u_i).
\label{eq:view}
\end{align}
Views share the central 4.096-ms interval containing the anomalies, without advancing scene time; confirmation uses correlated observations. Controllers share these \newhl{precomputed} scene--view IQ samples. Only acquired outputs become visible; failures supply no evidence and do not resample the scene. Hidden visibility and boxes support offline scoring. \revhl{Figure}~\ref{fig:iq-case} \revhl{illustrates the operator-facing request and report around the internal acquisition--feedback loop, alongside the acquired IQ and detector evidence.} \revhl{Catalog and filter details appear in \mbox{the supplement.}}

The frozen spectral-residual detector computes a log-power short-time Fourier transform with a 256-sample window and 128-sample hop, then subtracts each frequency bin's temporal median. The residual's 99.5th percentile is the view score. Calibration uses all 5,400 views from 900 normal primary-validation scenes; the 95th-percentile threshold is frozen at 10.34375 dB. \revhl{The 5\% false-positive target specifies the operating point on \mbox{calibration views.}}

\revhl{Confidence is an uncalibrated detector-margin score, computed as the sigmoid of the threshold margin divided by 2~dB.} For positive views, the thresholded connected component with greatest summed residual supplies a normalized time--frequency box; normal views lack valid boxes. Persistent events can attenuate their residual by entering the median background. With each view attempted at most once, confirmation depends on the number of obtainable positive or normal outputs, even under \mbox{shared perception.}

\subsubsection{\revhl{Baselines and Training Configuration}}
\label{sec:controller-setup}
\revhl{The main experiments use Qwen3-8B with thinking disabled \mbox{\cite{yang2025qwen3}}. Independently trained interface and planning controllers are specified in Section}~\ref{sec:control-decomposition}; \revhl{Section}~\ref{sec:backbone-diagnostic} \revhl{reports additional-backbone training diagnostics.} \revhl{Three fixed policies implement balanced, recall-priority, or efficient behavior. The \newhl{term frequency--inverse document frequency (TF--IDF)} router selects among seven hand-coded policies acting on changing public state, retaining the agent loop. Full-state TF--IDF predicts a Macro from \mbox{$(x,s_t)$}; the LLM predicts a Macro and reason code. Shared perception, execution, and reporting keep the surrounding \mbox{architecture fixed.}}

\revhl{Standard SFT learns from 20,000 next-action demonstrations under three original profiles and a four-call budget. Each training row pairs public state with teacher Macro/reason JSON, including stopping. Hard SFT adds 10,000 instruction--state pairs from 300 training scenes, crossing initialization, partial evidence, failure, and recovery with 30 paraphrase, composition, and failure-policy templates. The supplement records sampling quotas and label construction.} \revhl{Strategy and parsing share adapters. Training uses rank-16 LoRA, one epoch per stage, and a 1024-token limit; full hyperparameters and stage-specific learning rates appear in \mbox{the supplement.}}

\revhl{Hard DPO continues Hard SFT on 5,000 RF-proxy preference pairs. Offline scoring uses teacher rollouts and hidden labels, omitting audited compliance and applying a historical any-positive report rule. Full construction details appear in \mbox{the supplement.}}

\revhl{Full-state TF--IDF uses the same 30,000 SFT rows, with regularization selected on 8,100 pairs from 60 validation scenes. The router uses 33 deduplicated instructions and seven policy categories. Feature dimensions and tuning choices appear in \mbox{the supplement.}}

\revhl{Evaluation corrects two archived training conventions: failed views no longer count toward coverage, and the conditional call cap takes priority over retry. The archived adapters and full-state classifier retain their original training labels. Standard SFT also contains 3,142 examples above the 1024-token training limit (15.71\%); hard-state examples fit. The supplement details these runtime and \mbox{truncation differences.}}

\subsubsection{Execution Settings and Reproducibility}
\revhl{All experiments use greedy generation, thinking disabled, at most 4096 input tokens without truncation, and batches of up to four. Validation confirmed original-experiment batch consistency. Direct actions and plans undergo structural, view, resource, and token-limit checks; plans also reject repeats and excessive length. Output-error rates combine these failures; transport and service failures are \mbox{recorded separately.}}

%Supplementary local tests permute view indices at fixed physical bands and shift RF distributions; earlier experiments use different data or execution settings. Evaluation protocols, training summaries, and numerical results are provided in the \href{https://github.com/haozhangcn/rf_agent/tree/37fa8f6}{accompanying repository}. Full reproduction additionally requires the separately stored episode records, IQ data, and model files identified in \mbox{its manifests.}

\subsection{Evaluation Protocol}
\label{sec:protocol}
\subsubsection{Evaluation Suites and Data Separation}
Suites use nested prefixes of the deterministically shuffled Core manifest, fixed before inference without rebalancing (Table~\ref{tab:cohorts}). The changed-requirement subset contains 19 normal and 21 anomalous scenes; the supplement gives full \mbox{class counts.}

\begin{table*}[t]
\centering
\caption{\revhl{Evaluation Cohorts and Conditions}}
\label{tab:cohorts}
\small
\setlength{\tabcolsep}{4pt}
\begin{tabularx}{\textwidth}{@{}p{0.19\textwidth}p{0.17\textwidth}Xp{0.22\textwidth}@{}}
\toprule
Suite & Source scenes & Instructions, states, and attempt budgets & Cases or episodes\\
\midrule
Matched-state & First 30 Core & 5 snapshots; 11 instructions at all states, 2 with failure history; $B=4$ & 1,770 cases; 4,340 CF pairs\\
Core & First 120 Core & 13 instructions; $B=4$ & 1,560 episodes\\
Changed requirements & First 40 Core & 4 tasks; $B\in\{1,3,5\}$ & 480 episodes\\
\midrule
Control Core & Same 120 Core & 13 instructions; $B\in\{1,2,3,4\}$ & 6,240 episodes; 1,560 per budget\\
Control changed requirements & Same 40 Core & 4 tasks; $B\in\{1,3,5\}$ & 480 episodes; 160 per budget\\
\bottomrule
\end{tabularx}
\par\vspace{3pt}
\begin{minipage}{\linewidth}
\footnotesize\normalfont\raggedright
\revhl{The first three rows cover strategy/parsing; the last two cover interface/planning. Repeated source scenes stay grouped during resampling.}
\end{minipage}
\end{table*}

Training supplies demonstrations; validation calibrates perception and selects classifier regularization, without test scenes. The 13 original instructions comprise three intentionally seen controls, six unseen paraphrases, two compositions, and two failure policies. Standard SFT includes the controls; the hard-state corpus contains no test instructions verbatim, although related requirements occur in training. Four author-written changed-requirement prompts reuse Core scenes, testing task and failure changes within the Core RF distribution. Table~\ref{tab:contracts} lists the corresponding \mbox{task specifications.}

\begin{table}[t]
\centering
\caption{\revhl{Task Specification Parameters}}
\label{tab:contracts}
\small
\setlength{\tabcolsep}{3pt}
\renewcommand{\arraystretch}{1.12}
\begin{tabularx}{\columnwidth}{@{}Xcccc@{}}
\toprule
Task & $(k_+,k_-)$ & $f$ & $L$ & $(d_-,d_+)$\\
\midrule
Balanced, recall, efficient & $(1,1)$ & None & \NA & $(0,0)$\\
Two-positive confirmation & $(2,2)$ & None & \NA & $(1,0)$\\
Two-call normal cap & $(1,1)$ & Retry & $2$ & $(0,1)$\\
Recover after timeout & $(1,1)$ & Retry & \NA & $(0,0)$\\
Abort after timeout & $(1,1)$ & Abort & \NA & $(0,0)$\\
\midrule
Three-positive confirmation & $(3,2)$ & None & \NA & $(1,0)$\\
Confirmation with retry & $(2,2)$ & Retry & \NA & $(1,0)$\\
Confirmation with abort & $(2,2)$ & Abort & \NA & $(1,0)$\\
Three-call normal cap & $(1,1)$ & Retry & $3$ & $(0,1)$\\
\bottomrule
\end{tabularx}
\par\vspace{3pt}
\begin{minipage}{\columnwidth}
\footnotesize
$k_+$/$k_-$: positive/normal report thresholds; $f$: failure rule; $L$: call cap before any positive; $d_-$: mandatory stop after sufficient normal evidence; $d_+$: mandatory continuation after a positive. None adds no failure-specific obligation; \NA\ denotes no conditional cap. The first row also covers six instruction paraphrases.
\end{minipage}
\par\vspace{3pt}
\begin{minipage}{\linewidth}
\footnotesize\normalfont\raggedright
\revhl{The first five rows are original task families; the last four change requirements.}
\end{minipage}
\end{table}

\subsubsection{Interface and Planning Comparisons}
\label{sec:control-decomposition}
\paragraph{Hierarchical versus direct-action \newhl{(flat)} control}
\revhl{Two Qwen3-8B adapters share 20,000 paired public states, sample order, base initialization, and one-epoch training. Targets are a Macro or direct acquisition/stopping; supervised token counts differ. Both use seed 3407, rank-16 LoRA, and learning rate \mbox{$2\times10^{-4}$}. Resource checks and reporting remain shared across 13,440 \mbox{evaluation episodes.}}

\paragraph{Feedback versus initial-state planning}
\revhl{An open-loop adapter emits a fixed view plan from \mbox{$(x,s_1)$}. Its 18,000 episode-level examples include conflicting trajectory-derived plans for identical inputs. Feedback access, training unit, and target ambiguity therefore vary together. Fixed-\mbox{$K$} plans use initial geometry for each \mbox{$K=1,\ldots,B$}, without per-scene oracle selection. This campaign covers 30,480 episodes and independently evaluates its hierarchical controller.} \revhl{All new interface/planning examples fit the 2048-token training limit. The supplement gives full training counts, target-ambiguity statistics, and every \mbox{budget result.}}

\paragraph{Once-per-episode instruction parsing}
The SFT and DPO Macro adapters parse each episode's instruction once, without parser SFT or cross-episode caching. Outputs are the six fields of $g$ plus an optional total-attempt limit, \texttt{max\_actions}, which is null in all reference contracts. A predicted limit can further restrict acquisition within the environment budget. \revhl{In this comparison, a deterministic policy uses the predicted specification \mbox{$\hat g$} to select actions from changing public states; the shared reporter and auditor retain the reference \mbox{$g$}. The evaluated scope is instruction parsing for control with reference-specification reporting.} The comparison includes 24,780 matched decisions and 28,080 episodes with \mbox{shared baselines.}

\subsubsection{Matched-State Instruction Tests}
\label{sec:matched-protocol}
Each of 30 scenes supplies five shared snapshots at $B=4$: initialization, successful \textsc{Search}, and successful \textsc{Gather} on one path; forced first-\textsc{Search} failure and successful \textsc{Gather} on another view on a separate path. Eleven instructions apply to all 150 snapshots; two failure policies apply to the 60 failure-history snapshots, yielding \mbox{$11\times150+2\times60=1770$ cases.}

For applicable cases $\mathcal Q$ and prescribed teacher Macro $m^\star(x,s)$, reference-Macro accuracy is
\begin{equation}
A_{\mathrm M}=\frac{1}{|\mathcal Q|}
\sum_{(x,s)\in\mathcal Q}\indicator[\pi(x,s)=m^\star(x,s)].
\end{equation}
Let $\mathcal P$ contain unordered applicable instruction pairs at the same snapshot with different teacher labels. With $\hat m_i=\pi(x_i,s)$ and $m_i^\star=m^\star(x_i,s)$, counterfactual-pair accuracy is
\begin{equation}
A_{\mathrm{CF}}=\frac{1}{|\mathcal P|}
\sum_{(i,j,s)\in\mathcal P}
\indicator[\hat m_i=m_i^\star\ \land\ \hat m_j=m_j^\star].
\label{eq:cf}
\end{equation}
Both predictions must match their labels. Equal weighting of the 4,340 pairs favors snapshots with more conflicting labels. These metrics assess prescribed Macro choices. Physical control is evaluated through the executed trajectories, since different Macros can select the same view and compliant actions can differ from the teacher.

\subsubsection{Closed-Loop Failures and Task Feasibility}
\label{sec:closed-loop-protocol}
Episodes start with empty histories. Failure probability is 0.25 for the original conditional-budget and two failure-handling tasks, and zero otherwise. All four changed-requirement tasks use a random failure probability of 0.25 and force the second attempt to fail. The forced failure overrides the random outcome. SHA-256 of seed 37, scene identity, and zero-based attempt index supplies the random draw. At a fixed probability, controllers share failure sequences regardless of view, budget, or layout. Realized rates depend on the \mbox{attempts reached.}

These task and failure changes jointly restrict feasibility. Table~\ref{tab:feasibility} assumes successful nonforced attempts and sufficient obtainable detector outputs. At $B=1$, all confirmation tasks lack enough observations. At $B=3$, three-positive confirmation can succeed only on the 19 normal scenes. Confirmation with abort allows at most one success before mandatory termination, precluding a compliant binary report at every evaluated budget. Proposition 2 therefore gives $J\leq259/480\approx53.96\%$ across the 480 episodes: at least 221 cannot achieve joint success. All remain in the denominators. Compliance, resource limitation, and budget-resolved results distinguish policy violations from these protocol-imposed \mbox{evidence limits.}

\begin{table}[t]
\centering
\caption{\revhl{Evidence-Capacity Bounds Under Changed Requirements}}
\label{tab:feasibility}
\small
\setlength{\tabcolsep}{4pt}
\begin{tabularx}{\columnwidth}{@{}Xccc@{}}
\toprule
Task & $B=1$ & $B=3$ & $B=5$\\
\midrule
Three-positive confirmation & $0$ & $19/40$ & $1$\\
Confirmation with retry & $0$ & $1$ & $1$\\
Confirmation with abort & $0$ & $0$ & $0$\\
Three-call normal cap & $1$ & $1$ & $1$\\
\bottomrule
\end{tabularx}
\par\vspace{3pt}
\begin{minipage}{\linewidth}
\footnotesize\normalfont\raggedright
\revhl{Each cell is an upper bound on the joint-success rate over 40 scenes, assuming successful nonforced attempts and sufficient obtainable RF evidence.}
\end{minipage}
\end{table}

\subsubsection{Outcome Metrics and Paired Statistics}
\label{sec:outcome-metrics}
\revhl{All closed-loop comparisons report joint success \mbox{$J$} in \mbox{\eqref{eq:joint-success}} alongside compliance and \mbox{RF correctness.} Joint success counts their conjunction in each episode; it is not the product of the two aggregate rates.} Accuracy $\mathrm{Acc}$ is the mean of $R_e$ over all episodes; anomaly recall $\mathrm{Rec}$ and normal specificity $\mathrm{Spec}$ are the corresponding means over anomaly and normal scenes. Let $A_e$ and $B_e$ denote attempted acquisitions, including failures, and the initial effective budget of episode $e$. The balanced, recall-priority, and efficient profiles share hard obligations but express different RF preferences through
\begin{align}
U_{\mathrm{bal}}&=\mathrm{Acc},\nonumber\\
U_{\mathrm{rec}}&=0.8\,\mathrm{Rec}+0.2\,\mathrm{Spec},\nonumber\\
U_{\mathrm{eff}}&=0.8\,\mathrm{Acc}
 +0.2\,\mathbb E_e[\max(0,1-A_e/B_e)].
\label{eq:utility}
\end{align}
The weights are predefined benchmark choices. These utilities quantify RF preferences; compliance is assessed separately. A condition combines instruction, requested budget, and receiver layout. Each rate and the expectation in \eqref{eq:utility} are computed within a condition before equal-weight averaging of defined condition utilities; recall utility requires both scene classes. All four changed-requirement tasks use $U_{\mathrm{bal}}$, whose mean equals pooled accuracy for their equally sized conditions.

\revhl{Costs count RF attempts including failures, model requests, and input/output tokens. Error-terminated episodes retain incurred costs; lower acquisition counts can reflect interrupted execution. Local policies incur computation but no model requests. Training, service operations, and missing request usage \mbox{are excluded.}}

% Evidence-based abstention follows \eqref{eq:report}; controller errors force abstention as specified in Section~\ref{sec:execution}. Resource limitation and obligation exposure follow Section~\ref{sec:obligation-audit}. A discovery occurs when a successful view overlaps the anomaly in both time and frequency. We compute the discovery rate over anomalous episodes. Recovery counts failed attempts followed by later success; two failures preceding one success count twice. Retry compliance requires an attempt. RF scoring uses hidden truth; audits use public evidence.

We estimate paired 95\% confidence intervals with 2,000 percentile-bootstrap resamples and seed 913. Each scene is resampled together with all its conditions, using the same sampled scenes for both controllers; matched-state accuracies, pair accuracies, and condition-level utilities are recomputed in each replicate. The matched-state, changed-requirement, and Core analyses use 30, 40, and 120 source scenes, respectively. The intervals describe uncertainty across scenes for the fixed models, instructions, and failure design. They exclude variation across training seeds and new operator instructions and are not adjusted for \mbox{multiple comparisons.}

\begin{table*}[t]
\centering
\caption{\revhl{Instruction Agreement and Core Outcomes (96-Token Protocol)}}
\label{tab:formal-main}
\small
\setlength{\tabcolsep}{8pt}
\begin{tabular*}{\textwidth}{@{\extracolsep{\fill}}lrrrrrr@{}}
\toprule
Controller & Macro acc. & CF acc. & Core compliance & Core RF & \revhl{Core joint} & RF attempts\\
\midrule
Exact teacher$^*$ & 100.00 & 100.00 & 100.00 & 73.27 & \revhl{73.27} & 2.835\\
Contract reference$^*$ & 49.04 & 8.25 & 100.00 & 59.10 & \revhl{59.10} & 1.189\\
\midrule
Fixed balanced & 67.23 & 0.00 & 80.96 & 75.45 & \revhl{63.33} & 3.116\\
Fixed recall & 63.33 & 0.00 & 85.38 & \textbf{77.88} & \revhl{66.92} & 4.000\\
Fixed efficient & 61.64 & 0.00 & 89.36 & 69.87 & \revhl{64.17} & 1.920\\
TF--IDF router & 94.69 & 89.06 & \textbf{100.00} & 72.82 & \textbf{\revhl{72.82}} & 2.695\\
Full-state TF--IDF & 80.73 & 53.43 & 93.14 & 72.95 & \revhl{69.17} & 2.712\\
Standard SFT & 66.55 & 24.03 & 86.41 & 72.50 & \revhl{65.13} & 2.753\\
\midrule
\textbf{\revhl{RF-Agent}} & \textbf{98.31} & \textbf{96.22} & 99.17 & 72.76 & \revhl{72.12} & 2.708\\
\bottomrule
\end{tabular*}
\par\vspace{3pt}
\begin{minipage}{\linewidth}
\footnotesize\normalfont\raggedright
\revhl{Rates are percentages: 1,770 matched cases, 4,340 CF pairs, and 1,560 Core episodes per controller. Joint requires compliance and RF correctness; RF attempts are episode means and include failures. \newhl{RF-Agent uses Hard SFT in this campaign.} Bold: best non-reference rates. Asterisks: specification-aware references.}
\end{minipage}
\end{table*}

\begin{figure*}[!t]
\centering
\includegraphics[width=\textwidth,trim=0 3pt 0 7pt,clip]{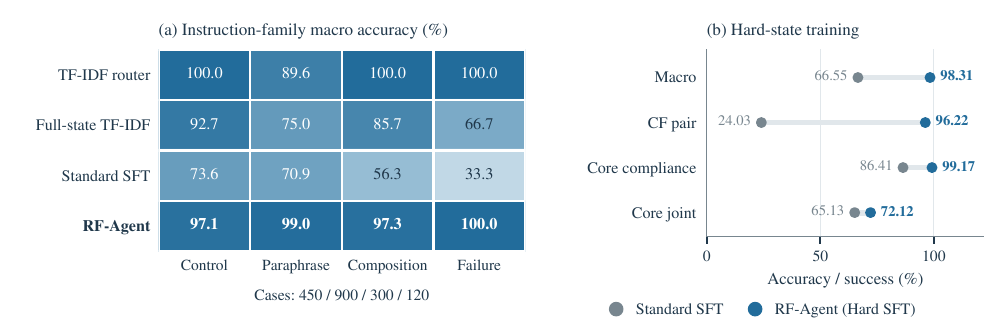}
\caption{\revhl{Instruction responsiveness and hard-state training (96-token campaign).}}
\label{fig:matched-diagnostics}
\end{figure*}

\subsection{Task Completion over Full Sensing Trajectories}
\label{sec:results}
\revhl{We evaluate RF-Agent through task completion, instruction responsiveness, execution reliability, and feedback-dependent acquisition. RF-Agent denotes the hierarchical controller with the training configuration of each campaign. The main 96-token experiment uses Hard SFT; the interface and planning studies independently train hierarchical controllers. Table}~\ref{tab:campaigns} \revhl{separates their training and evaluation settings. Macro and CF measure supervisory agreement; compliance, RF correctness, and joint success measure \mbox{complete trajectories.}}

\revhl{Table}~\ref{tab:formal-main} \revhl{connects supervisory decisions to Core task outcomes. Hard-state training raises joint success from 65.13\% to 72.12\% (+6.99 percentage points): 1,125 versus 1,016 of the same 1,560 episodes are both compliant and RF-correct. Compliance increases from 86.41\% to 99.17\% (+12.76 points; paired 95\% CI: 11.99--13.59), while RF correctness is 72.76\% versus 72.50\%. The identity \mbox{$J=\mathrm{Acc}-\ell_{\rm cmp}$}, where \mbox{$\ell_{\rm cmp}$} is \newhl{the fraction of all episodes that are RF-correct but noncompliant}, decomposes the joint gain into 0.26 points of RF accuracy and 6.73 points of reduced compliance loss. RF-correct but noncompliant outcomes fall from 115 to 10; these are aggregate counts, not paired trajectory transitions. Mean acquisitions decrease from 2.753 to 2.708.}

\revhl{The lightweight controls distinguish RF detection from instruction-conditioned control. Fixed recall achieves 77.88\% RF correctness but 66.92\% joint success, because RF evidence alone does not establish compliance. TF--IDF routing achieves 72.82\% joint success with 100\% compliance on these original task templates; RF-Agent achieves 72.12\% while leading matched-state agreement. The router selects one of seven policies from the instruction; the selected hand-coded policy then acts on updated states. RF-Agent predicts each Macro directly from the instruction and current state. These results distinguish agreement with instruction-dependent supervisory choices from successful trajectory execution, with lightweight routing providing a strong task-completion baseline. The complete baseline comparison, condition-averaged utility, and paired intervals are retained in \mbox{the supplement.}}

\subsection{Instruction Responsiveness and Supervisory Training}
\revhl{RF-Agent achieves 98.31\% Macro and 96.22\% CF accuracy, compared with standard SFT's 66.55\% and 24.03\% (Fig.}~\ref{fig:matched-diagnostics}\revhl{). CF requires both decisions to match distinct instruction-dependent labels at an identical RF state. Hard-state training therefore connects a 72.19-point CF gain with the 6.99-point Core joint-success improvement, rather than relying on aggregate RF \mbox{accuracy alone.}}

\begin{figure}[t]
\centering
\includegraphics[width=\columnwidth]{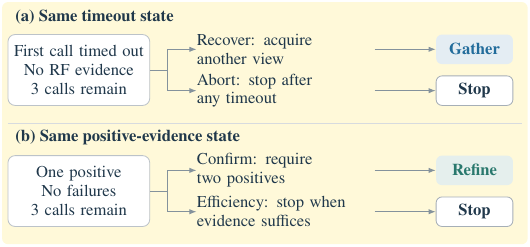}
\caption{\casehl{Recorded RF-Agent decisions (Qwen3-8B Hard SFT, 96-token evaluation). Within each pair, only the instruction changes. Instructions are condensed; all four outputs match teacher Macros. Panels use distinct snapshots; agreement does not establish full-trajectory compliance.}}
\label{fig:recorded-decisions}
\end{figure}

\casehl{Figure}~\ref{fig:recorded-decisions} \casehl{illustrates actual instruction-dependent decisions. In (a), Search at the -0.5-MHz view times out. Identical failure feedback elicits Gather under recovery and Stop under abort. In (b), that view yields a positive score of 28.03 dB, above the 10.34375-dB threshold. Two-positive confirmation elicits Refine; efficiency elicits Stop because one positive suffices for that task. Within each pair, evidence, history, available views, and budget are identical. These selected single-step predictions illustrate the aggregate Macro/CF results; they do not establish subsequent RF correctness or full-trajectory compliance.}

\revhl{Relative to TF--IDF routing, RF-Agent gains 3.62 Macro and 7.17 CF points. One recall-priority paraphrase accounts for all 94 TF--IDF errors, with 37.33\% accuracy versus RF-Agent's 94.00\%. The TF--IDF router reaches 100\% on compositions, versus 97.33\% for RF-Agent. Full-state TF--IDF, which directly predicts a Macro from instruction and state, reaches 80.73\% Macro and 53.43\% CF accuracy. These comparisons locate the learned supervisor's contribution in instruction-dependent action selection across the evaluated state and \mbox{language variations.}}

\begin{figure*}[!t]
\centering
\includegraphics[width=0.94\textwidth,trim=0 3pt 0 7pt,clip]{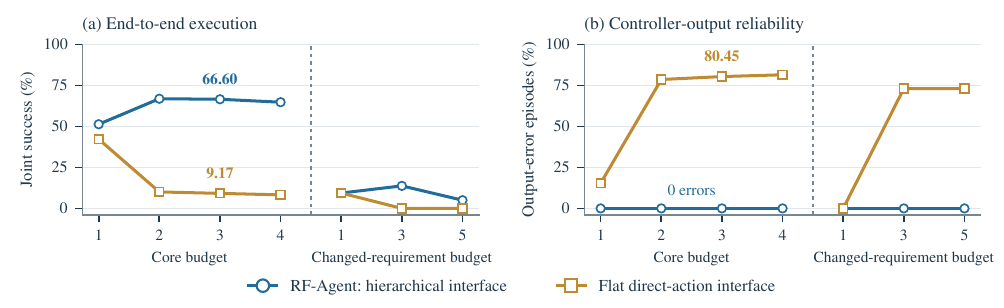}
\caption{\revhl{Interface campaign. Each Core/changed-requirement budget point contains 1,560/160 episodes, including output-error episodes.}}
\label{fig:execution-reliability}
\end{figure*}

\revhl{RF-proxy preference optimization provides a training diagnostic. Continuing Hard SFT with Hard DPO reduces Macro/CF accuracy to 88.93\%/72.93\% and Core compliance/joint success to 96.09\%/68.78\%. The preference construction omits audited compliance and uses \newhl{an any-positive anomaly rule}. This RF-proxy preference continuation does not preserve the instruction compliance achieved by Hard SFT; complete DPO scores and construction details appear in \mbox{the supplement.}}

\subsection{Architectural Execution Reliability}
\revhl{The paired-state interface study evaluates the practical consequence of delegating receiver-action construction to an executor. At Core budget three, hierarchical RF-Agent reaches 66.60\% joint success versus 9.17\% for flat control (+57.44 points; paired 95\% CI: 50.83--63.46). Flat control records 80.45\% controller-output errors, while the hierarchical controller records none (Fig.}~\ref{fig:execution-reliability}\revhl{). Proposition 2 bounds flat joint success by its 19.55\% error-free fraction (305/1,560); its observed joint successes number 143. Among error-free episodes, flat joint success is 46.89\%. The comparison identifies output validity as a task-success bottleneck addressed by the hierarchical interface: its four labels delegate resource-valid action construction to deterministic execution. Conditional rates describe different controller-selected subsets, not a matched \mbox{causal comparison.}}

\begin{figure*}[!t]
\centering
\includegraphics[width=0.94\textwidth,trim=0 3pt 0 7pt,clip]{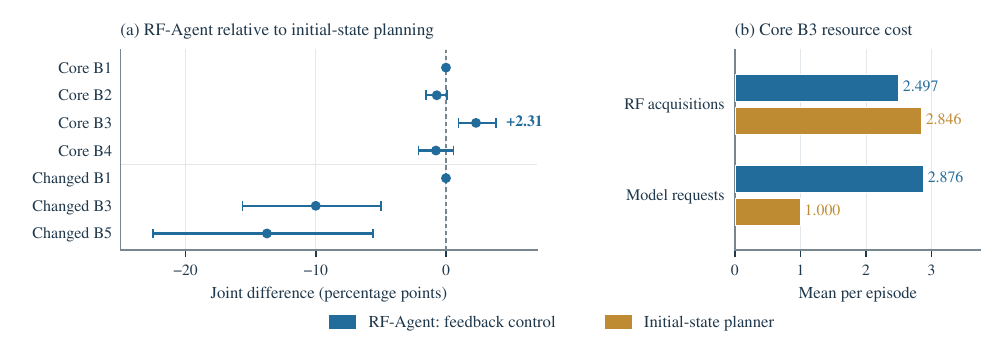}
\caption{\revhl{Planning campaign (1,560/160 episodes per Core/changed-requirement budget point). Panel (a) shows RF-Agent-minus-planner differences with paired 95\% scene-bootstrap intervals.}}
\label{fig:feedback-cost}
\end{figure*}

\revhl{Across Core budgets two to four, hierarchical joint success is 66.92\%, 66.60\%, and 64.81\%, versus flat control's 10.06\%, 9.17\%, and 8.27\%; flat output errors range from 78.65\% to 81.54\%. Under changed requirements, hierarchical joint success reaches 13.75\% and 5.00\% at budgets three and five, versus 0\% for flat control. Both reach 9.38\% at budget one. The measured effect covers the complete interface and executor: output formats, supervised tokens, and training targets differ. Error counts combine formatting, view-selection, and action-validity failures, so lower flat acquisition counts reflect \mbox{interrupted execution.}}

\subsection{Feedback and Acquisition--Inference Cost}
\revhl{In the independent planning campaign, RF-Agent uses current observations and failures to revise acquisition and stopping. At Core budget three, it improves joint success over an initial-state planner from 64.29\% to 66.60\% (+2.31 points; paired 95\% CI: 0.96--3.85) while reducing RF acquisitions from 2.846 to 2.497 (12.3\%). Compliance is 90.13\% versus 86.35\%, with RF correctness of 71.35\% versus 73.33\%. Figure}~\ref{fig:feedback-cost} \revhl{places this operating point alongside every evaluated budget and its \mbox{acquisition--inference cost.}}

\revhl{\newhl{At Core budget three, the feedback controller averages 2.876 model requests and 2.497 RF acquisitions per episode}, consistent with Proposition 1. The 591 excess requests over 3,895 acquisitions are terminal Stop decisions across 1,560 episodes. \newhl{The planner uses one request per episode. Across these 1,560 episodes, the feedback controller uses 3,106,199 tokens and the planner uses 1,031,880.} Costs cover completed requests and exclude training, unavailable service-failure usage, monetary cost, and latency.}

\revhl{Fixed two-call and three-call acquisition attain 65.00\% and 65.19\% joint success with no model requests; RF-Agent's corresponding differences of 1.60 and 1.41 points have intervals spanning zero. At Core budgets two and four, RF-Agent-minus-planner differences are -0.71 and -0.77 points, also with intervals spanning zero. The planner leads by 10.00 and 13.75 points under changed requirements at budgets three and five. These complete-controller comparisons retain differences in feedback, output formats, training units, and initial-plan \mbox{target ambiguity.}}

\begin{figure*}[!t]
\centering
\includegraphics[width=0.94\textwidth,trim=0 3pt 0 7pt,clip]{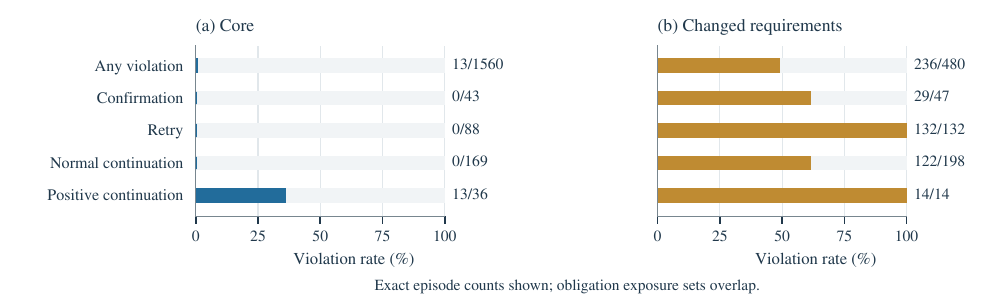}
\caption{\revhl{Continuation audits (96-token campaign). Overall violations use all episodes; obligation-specific rates use feasible continuation exposures after an earlier attempt, excluding resource-blocked cases.}}
\label{fig:constraint-diagnostics}
\end{figure*}

\paragraph{\revhl{Acquisition--inference cost threshold}}
\revhl{Assume constant costs \mbox{$c_{\rm RF}\geq0$} per RF attempt and \mbox{$c_{\rm LLM}>0$} per model request, in common units. For mean episode cost \mbox{$\mathcal C=c_{\rm RF}\bar A+c_{\rm LLM}\bar Q$}, Core budget-three feedback-minus-planner cost is}
\begin{equation}
\text{\revmath{\Delta\mathcal C\approx-0.349c_{\rm RF}+1.876c_{\rm LLM}.}}
\label{eq:feedback-cost-threshold}
\end{equation}
\revhl{Feedback therefore costs less when \mbox{$c_{\rm RF}/c_{\rm LLM}>5.38$}, alongside its observed 2.31-point joint-success gain. This budget-specific threshold uses rounded means, not measured prices or latency. Variable request costs require token- or \mbox{serving-cost accounting.}}

\revhl{A separate 256-token parsing diagnostic examines where instruction interpretation enters the loop. Per-step supervision reaches 72.05\% Core joint success versus 57.76\% for parsing once and applying a deterministic feedback policy (+14.29 points; paired 95\% CI: 8.72--20.32). Parsing once reduces requests by 69.64\% and tokens by 71.36\%. Both variants use feedback, and the Macro-trained adapters have no parser-specific SFT. Even with the reference specification, the deterministic policy reaches 59.10\% joint success, showing that the observed gap also reflects downstream acquisition and stopping decisions. The supplement reports parser agreement, full costs, and \mbox{changed-requirement results.}}

\begin{table}[!ht]
\centering
\caption{\revhl{Cross-Backbone Hard-SFT Agreement}}
\label{tab:backbone-diagnostic}
\small
\setlength{\tabcolsep}{3pt}
\renewcommand{\arraystretch}{1.12}
\begin{tabular*}{\columnwidth}{@{\extracolsep{\fill}}lcrr@{}}
\toprule
Backbone & \revhl{\mbox{$n_{\mathrm{seed}}$}} & Macro & CF pair\\
\midrule
Qwen3-8B & 3 & $98.76\pm1.08$ & $97.47\pm2.30$\\
Llama-3.1-8B & 3 & $95.97\pm1.37$ & $91.08\pm4.61$\\
Mistral-Nemo-12B & 1 & $99.94$ & $99.84$\\
\bottomrule
\end{tabular*}
\par\vspace{3pt}
\begin{minipage}{\linewidth}
\footnotesize\normalfont\raggedright
\revhl{Archived matched-state protocol. Accuracies are percentages; \mbox{$n_{\mathrm{seed}}$} counts seeds. Three-seed entries report mean \mbox{$\pm$} sample standard deviation.}
\end{minipage}
\end{table}

\subsection{Constraint Fulfillment and Evidence Feasibility}
\revhl{Trajectory audits connect aggregate scores to the agent's decisions. On Core, all 13 RF-Agent violations are premature stops among 36 episodes requiring and permitting \newhl{continued search after a positive detection under \mbox{$d_+=1$}}. It records no violations in the 43, 88, and 169 episodes exposed to positive-confirmation \newhl{(\mbox{$0<p_t<k_+$})}, recovery, and normal-evidence continuation obligations, respectively, after an earlier attempt (Fig.}~\ref{fig:constraint-diagnostics}\revhl{). These counts distinguish when an obligation actually becomes actionable from its presence in \mbox{an instruction.}}

\revhl{Under changed requirements, the audits identify recovery and continued evidence collection as the main failure modes: RF-Agent violates the retry obligation in all 132 episodes with a feasible retry exposure and the normal-evidence continuation obligation in 122 of 198 exposed episodes. These failures mark a transfer boundary: strong instruction responsiveness on the matched-state suite does not ensure compliance under changed task and failure requirements. Supplementary Table S2 reports the complete controller comparison; Fig.}~\ref{fig:constraint-diagnostics} \revhl{retains the associated \mbox{obligation diagnostics.}}

\revhl{Feasibility explains a separate component of task outcomes. Under confirmation with abort, the forced second-attempt failure prevents a compliant binary report at every evaluated budget. RF-Agent nevertheless reaches 65.00\% compliance on that task, versus 33.33\% and 43.33\% for TF--IDF routing and full-state TF--IDF. Across the entire suite, the optimistic joint-success ceiling is 53.96\% (Table}~\ref{tab:feasibility}\revhl{); the contract reference's 21.67\% is an achieved outcome rather than that ceiling. All infeasible episodes remain in aggregate denominators. Full task-by-budget results and resource-limitation diagnostics appear in \mbox{the supplement.}}

\subsection{Cross-Backbone Supervisory Training}
\label{sec:backbone-diagnostic}
\revhl{Table}~\ref{tab:backbone-diagnostic} \revhl{extends supervisory training to three model families, each using 20,000 standard-SFT and 10,000 hard-state examples; \newhl{reported Macro accuracy exceeds 95\% for each backbone}. The archived protocol contains 1,770 cases and 4,364 CF pairs (4,340 currently). Qwen and Llama each share a backbone-specific initialization across three continuation seeds; Mistral has one run. These are supervisory-agreement diagnostics; complete-agent campaigns use Qwen3-8B.}

\section{Conclusion}
\label{sec:conclusion}
\revhl{RF-Agent integrates instruction-responsive supervision, episode memory, deterministic execution, and perception feedback for active spectrum sensing. Execution bounds characterize supervisory requests and history-token volume; success and state-level bounds distinguish output reliability from evidence capacity. Hard-state training raises Core joint success \newhl{by 6.99 points (65.13\% to 72.12\%), including 6.73 points from reducing the fraction of all episodes that are RF-correct but noncompliant.} \newhl{Interface and planning comparisons quantify execution reliability and acquisition--inference tradeoffs, including a budget-specific cost threshold.} \newhl{ActiveRF-AD explains task completion beyond RF correctness through instruction interventions, trajectory audits, and feasibility bounds.} Future work includes matched closed-loop backbone studies, dedicated parsing training, broader requirements, and \mbox{receiver hardware.}}

% Balance the final bibliography columns after clean-text reflow.
\IEEEtriggeratref{3}
\IEEEtriggercmd{\balance}
\bibliographystyle{IEEEtran}
\bibliography{references_0930}
\end{document}